\documentstyle[prd,aps]{revtex}

\newcommand{\be}{\begin{equation}}
\newcommand{\ee}{\end{equation}}
\newcommand{\bea}{\begin{eqnarray}}
\newcommand{\eea}{\end{eqnarray}}
\newcommand{\bdm}{\begin{displaymath}}
\newcommand{\edm}{\end{displaymath}}
\newcommand{\ul}{\underline}

\newcommand{\diff}{d}

\newcommand{\codtil}{\tilde{\diff}^{\dagger}}
\newcommand{\p}{\partial}

\newcommand{\const}{\mbox{const.}}

\newcommand{\vp}{\varphi}
\newcommand{\vt}{\vartheta}

\newcommand{\gtens}{\mbox{\boldmath $g$}}

\newcommand{\gtiltens}{\tilde{\gtens}}
\newcommand{\ghattens}{\hat{\gtens}}
\newcommand{\gtil}{\tilde{g}}
\newcommand{\ghat}{\hat{g}}
\newcommand{\gbar}{\bar{g}}
\newcommand{\laptil}{\tilde{\Delta}}
\newcommand{\nabtil}{\tilde{\nabla}}
\newcommand{\asttil}{\tilde{\ast}}
\newcommand{\epstil}{\tilde{\epsilon}}

\newcommand{\nabhat}{\hat{\nabla}}
\newcommand{\asthat}{\hat{\ast}}
\newcommand{\epshat}{\hat{\epsilon}}

\newcommand{\fourfold}{(M,\gtens)}

\begin{document}
%%%%%%%%%%%%%%%%%%%%%%%%%%%%%%%

\title{Gauge-invariant perturbations of Schwarzschild black holes in
horizon-penetrating coordinates}

\author{Olivier Sarbach$^{1}$\thanks{sarbach@gravity.phys.psu.edu} \and Manuel
Tiglio$^{1,2}$\thanks{tiglio@gravity.phys.psu.edu} }

\address{1. Center for Gravitational Physics and Geometry,
Department of Physics,\\ The
Pennsylvania State University, University Park, PA
16802.}
\address{2. Department of Astronomy and Astrophysics,
The Pennsylvania State University,\\
University Park, PA 16802.}
\maketitle

\begin{abstract}
We derive a geometrical version of the Regge-Wheeler and Zerilli
equations, which allows us to study gravitational perturbations on an
arbitrary spherically symmetric slicing of a Schwarzschild black
hole. We explain how to obtain the gauge-invariant part of the metric
perturbations from the amplitudes obeying our generalized
Regge-Wheeler and Zerilli equations and vice-versa. We also give a
general expression for the radiated energy at infinity, and establish
the relation between our geometrical equations and the Teukolsky
formalism. The results presented in this paper are expected to be
useful for the close-limit approximation to black hole collisions, for
the Cauchy perturbative matching problem, and for the study of
isolated horizons.
\end{abstract}

%%%%%%%%%%%%%%%%%%%%%%%%%%%%%%%%%%%%%%%%%%%%%%%%%%%%%%%%%%%
\section{Introduction}
%%%%%%%%%%%%%%%%%%%%%%%%%%%%%%%%%%%%%%%%%%%%%%%%%%%%%%%%%%%

In the last decade, perturbation theory for black holes has played, in
several different ways, a key role in numerical and computational
relativity.  Already in the seventies it proved to be a very valuable
tool to predict gravitational waveforms from processes such as a
particle falling towards a black hole. Since the early nineties, due
to the network of interferometric gravitational wave detectors in
construction, there has been renewed interest in predicting waveforms
for strong sources of gravitational waves such as black hole
collisions. In particular, the first predictions using perturbation
theory in this new era have been quite striking \cite{pp}.  Some of
the applications of perturbation theory in recent years involved
computing the evolution for different conformally flat initial data
describing black holes in the close limit in order to predict radiated
energy and angular momentum \cite{insp_cl}, to provide both analytical
understanding and benchmarking of full numerical results \cite{bench},
or to quantify the amount of spurious radiation in conformally flat
initial data \cite{spinning} (see \cite{cl_review} for a general
review). The usual Regge-Wheeler (RW) - Zerilli \cite{RW},
\cite{Zerilli} and Teukolsky \cite{Teukolsky} formalisms have also
been extended to second order \cite{second}, a necessary step in
providing first-order perturbations with their own ``error
bars''\cite{e_bars}. Other recent approaches use black hole
perturbations to extend the computational domain in numerical
simulations to the radiative zone via Cauchy-perturbative matching
\cite{cpm}, or concentrate full numerical resources in the nonlinear
regime and let perturbation theory take over in the late stage of
black hole collisions \cite{lazarus}.

All of the applications just mentioned, though diverse, have a common
feature: they are limited to perturbations of Schwarzschild black
holes in Schwarzschild coordinates, and Kerr black holes in
Boyer-Lindquist coordinates. The reason for this is that, until very
recently, most of the initial data typically used in numerical
relativity were for maximal slicing, and thus reduced, in the various
regimes where perturbation is used (far region, late times, initially
close black holes, etc), precisely to the Schwarzschild and Kerr
spacetimes in Schwarzschild and Boyer-Lindquist coordinates,
respectively. In recent years, however, work was started on
Kerr-Schild-type initial data \cite{ks_idata}, which are not
maximal. Part of the motivation for introducing this new kind of
initial data is to avoid the typical grid stretching that maximal
slicings produce near the event horizon\footnote{One does not have to
move from maximal slicing to get rid of spurious radiation; it is
enough to use initial data that is conformally Kerr, instead of the
more usual conformally flat \cite{c_kerr}.}, a stretching that
eventually causes numerical simulations to crash \footnote{There is
some new evidence that these crashes can be avoided by excising the
singularity and appropriately choosing the shift vector
\cite{alc}.}. One is then faced with the fact that in order to
accommodate these new initial data, either for the close-limit
approximation or for Cauchy-perturbative matching, a formalism is
needed that allows perturbations in more general slicings than
Schwarzschild and Boyer-Lindquist.

Another important motivation for having such a formalism in place is
to study the recently developed {\it isolated-horizon} formalism
\cite{isolated} in the perturbative regime. For such studies, one
needs to be able to analyze a neighborhood of the background horizon,
which necessitates the use of horizon-penetrating coordinates.

The two most used approaches to black hole perturbations have been the
RW-Zerilli and the Teukolsky ones. Each of these methods has its own
advantages and limitations: The Teukolsky formalism can be used for
rotating black holes, but one cannot obtain the whole perturbed
geometry but, rather, $\Psi _4$ or $\Psi _0$ (this is enough to
compute radiation, though)\footnote{Actually, one can construct
solutions to the linearized vacuum equations from a potential (which
is not $\Psi _4$) that satisfies the Teukolsky equation
\cite{wald}. This approach is very interesting but there are some
issues that still have to be worked out before it can be
implemented. For example, how to give initial data to the
corresponding potential (in particular, can one obtain any linear
vacuum perturbation of the Kerr spacetime from some potential?), how
to construct gauge invariants and extract radiation, etc.}. The RW -
Zerilli technique, on the other hand, provides the whole perturbed
metric, but is limited to non rotating black holes.

The Teukolsky equation, in its original formulation, can, in fact, be
used to describe perturbations around any Petrov type-D background,
without relying on a particular choice of coordinates. Work has
started very recently on the application of this to Kerr-Schild black
hole perturbations \cite{ks_teu}.

This paper, in turn, develops an appropriate extension of the
RW-Zerilli formalism to perturbations of a Schwarzschild black holes
in arbitrary spherically symmetric coordinates. One can imagine a huge
variety of applications of such an extension; here we have concentrated
on the aspects of the formalism that we need in order to proceed with
our main motivations.  In order to generalize the RW-Zerilli
formalism, we start from a perturbation formalism introduced by
Gerlach and Sengupta \cite{GS} and derive two master equations which
hold in any spherically symmetric coordinates of the background, but
reduce to the equations obtained by Regge-Wheeler and Zerilli if one
uses the standard Schwarzschild coordinates.

Our approach is organized as follows. In section II we present the
basic formalism that decouples the field equations into the
generalized RW and Zerilli ones. The special cases with total angular
momentum $l=1$ and $l=0$ are treated carefully.  In section III we
work out a relation needed for Cauchy-perturbative matching, namely,
the one between the RW and Zerilli functions and the ADM
quantities. In section IV we establish the relation between the
present formalism and the Teukolsky one, a relation that is desirable
not only to compute the radiated energy and make contact with
\cite{ks_teu}, but also from a conceptual point of view.  Finally, in
section V we comment on the properties of the RW and Zerilli
equations, and on a numerical code that we have written to solve them.
In order to establish the contact between the abstract formalism in
the body of this paper and more direct applications, we give some
explicit expressions in appendix A. Finally, in appendix B, we
summarize some properties of spin-weighted spherical harmonics which
are needed in section IV.

%%%%%%%%%%%%%%%%%%%%%%%%%%%%%%%%%%%%%%%%%%%%%%%%%%%%%%%%%%%
\section{The generalized RW and Zerilli equation}
\label{Sect-2}
%%%%%%%%%%%%%%%%%%%%%%%%%%%%%%%%%%%%%%%%%%%%%%%%%%%%%%%%%%%

In what follows, we assume that the background spacetime $\fourfold$
can be represented as a product of $\tilde{M} = M / \mbox{SO(3)}$
and $S^2$ with metric
\be
\gtens = \gtil_{ab}\,\diff x^a \diff x^b + r^2\,\ghat_{AB}\,\diff x^A \diff x^B \,.
\label{Eq-Metric}
\ee
Here $\ghattens = \diff\Omega^2$ is the standard metric on $S^2$, and
$\gtiltens$ and $r$ denote the metric tensor and a positive function,
respectively, defined on the two-dimensional pseudo-Riemannian orbit
space $\tilde{M}$. In what follows, lower-case Latin indices refer to
coordinates on $(\tilde{M}, \gtiltens)$, while capital Latin indices
refer to the coordinates $\vt$ and $\vp$ on $(S^2, \ghattens)$.
Below, we will derive perturbation equations which do not depend
explicitly on the background metric coefficients.  In fact, we will
only use the background equations which are given by the components of
the Einstein tensor

\bea
G_{ab} &=& - \frac{2}{r} \nabtil_a \nabtil_b r + \frac{1}{r^2} \left( 2r\laptil r + N - 1 \right) \gtil_{ab} \, ,
\label{Eq-Gab}\\
G_{AB} &=& \left( r\laptil r - r^2\tilde{\kappa} \right) \ghat_{AB} \, , \nonumber\\
G_{Ab} &=& 0. \nonumber
\eea
Here, $N = \gtiltens(\diff r,\diff r)$, and $\tilde{\kappa}$
denotes the Gauss curvature of the metric $\gtiltens$.
A coordinate-invariant definition of the ADM mass is given by
\bdm
M = \frac{r}{2}(1 - N).
\edm

We can see by inspection that this $M$ is the mass if Schwarzschild
coordinates are used; on the other hand, $M$ is defined in terms of
scalars on $\tilde{M}$, and therefore $M$ represents the ADM mass in
any coordinates on $\tilde{M}$.  Note that in a vacuum spacetime,
equation (\ref{Eq-Gab}) implies that $0 = r^2 (G_{ab}\nabtil^b r -
G^b_{\;\; b}\nabtil_a r) = \nabtil_a[r(1-N)]$ which shows that $M$ is
constant.

Since the background is spherically symmetric, it is convenient to
expand the perturbed metric in spherical harmonics: \bea \delta
g_{ab} &=& H_{ab} Y, \nonumber \\ \delta g_{Ab} &=& Q_b \nabhat_A
Y + h_b S_A, \nonumber \\ \delta g_{AB} &=& K g_{AB} Y + F
\nabhat_A\nabhat_B Y + 2k \nabhat_{(A} S_{B)}, \nonumber \eea
where $H_{ab}$ denotes a tensor field, $Q_b$ and $h_b$ vector
fields, and $K$, $F$ and $k$ scalar fields on $\tilde{M}$.  Here,
$Y \equiv Y^{l m}$ are the standard spherical harmonics, and $S_A
= (\asthat\diff Y)_A = \epshat^B_{\; A} \nabhat_B Y$ and
$2\nabhat_{(A} S_{B)} \equiv \nabhat_A S_B + \nabhat_B S_A$ form a
basis of odd-parity vector fields and symmetric tensor fields,
respectively, on $S^2$ (See Appendix D of Ref. \cite{SHB-Odd} for
more details on spherical tensor harmonics.).  We suppress the
indices $l m$ and the sum over these indices since the modes
belonging to different pairs of $l m$ decouple in the perturbation
equation. The $Y^{l m}$ are normalized with respect to the
standard metric $\ghattens$ on $S^2$, an exception being the cases
$l=0$ and $l=1$: There, we choose the normalization such that
$Y^{0 0}=1$, and $\int_{S^2}Y^{1 m} \bar{Y}^{1m} d\Omega = 4\pi
/3$.

In what follows, it will also be convenient to use a coordinate-free
notation for differential forms on $(\tilde{M}, \gtiltens)$: $\asttil$
and $\codtil \equiv \asttil\diff\asttil$ denote the Hodge dual and the
co-differential operator, respectively, with respect to
$\gtiltens$. That is,

\begin{eqnarray*}
\asttil u_a dx^a & = & \tilde{\epsilon}_{ab} u^a dx^b, \\
\asthat v_a dx^A & = & \hat{\epsilon}_{AB} v^A dx^B,
\end{eqnarray*}
where $\tilde{\epsilon}$ and $\hat{\epsilon}$ denote the standard
volume elements in $(\tilde{M},\tilde{g})$ and $(S^2,\hat{g})$,
respectively\footnote{For $\tilde{\epsilon}$ we need to provide an
orientation in $\tilde{M}$; if $t$ and $x$ are timelike and spacelike
coordinates, respectively, we choose $\tilde{\epsilon}_{tx} =
|\tilde{g}|^{1/2}$.}.  For example, we have

\bdm \codtil h = -\nabtil^a h_a, \;\;\;
(\codtil\diff h)_a = 2\nabtil^b \nabtil_{[a} h_{b]}.
\edm
A further simplification comes from the fact that a spherically symmetric
metric is invariant under parity transformation $\ul{x} \mapsto -\ul{x}$.
As a consequence, the above defined amplitudes decouple into two sets,
one set transforming like $Y$ (called {\it scalar perturbations\/}
or {\it even-parity perturbations\/}) and the other set transforming like
$S = \asthat\diff Y$ (called {\it vector perturbations\/} or
{\it odd-parity perturbations\/}) under parity transformations.
In this sense, the amplitudes $H_{ab}$, $Q_b$, $K$ and $F$ have even parity
while the amplitudes $h_b$ and $k$ have odd parity.

\subsection{The odd-parity sector}

We start with the simpler case of the odd-parity sector. The
perturbations of $g_{\mu\nu}$ are parameterized in terms of a scalar
field $k$ and a one-form $h = h_a \diff x^a$,

\be \delta g_{ab} = 0, \;\;\;
\delta g_{Ab} = h_b S_A, \;\;\;
\delta g_{AB} = 2k \nabhat_{(A} S_{B)} ,
\label{Eq-OddDeltag}
\ee
where $k$ and $h_a$ depend on the coordinates $x^b$ only.
Note that for $l=1$, $\nabhat_{(A} S_{B)}$ vanishes and
$k$ is not present.
For $l=0$, $S_A=0$ and there are no gravitational perturbations.

%%%%%%%%%%%%%%%%%%%%%%%%%%%%%%%%%%%%%%%%%%
\subsubsection{Coordinate-invariant amplitudes}
%%%%%%%%%%%%%%%%%%%%%%%%%%%%%%%%%%%%%%%%%%
A vector field $X = X^\mu \p_\mu$ generating an infinitesimal
coordinate transformation with odd parity is determined by a
function $f(x^b)$, where
$$
X^a = 0 , \;\;\;
X^A = \frac{f}{r^2} \, \ghat^{AB} S_B.
$$

Using the fact that to linear order, $\delta g_{\mu\nu}$
transforms like the Lie derivative of the background metric with
respect to $X$, we find the following transformations

\begin{eqnarray*}
h &\mapsto& h + r^2 \diff\left( \frac{f}{r^2} \right), \nonumber\\
k &\mapsto& k + f.
\end{eqnarray*}
Note that one can choose a gauge in which $k=0$. This gauge, which
is usually called the RW gauge, is unique.

For $l\geq 2$, one can construct the
coordinate-invariant one-form $$
h^{(inv)} \equiv h - r^2 \diff\left( \frac{k}{r^2} \right).
$$
For $l=1$, we will see that only the invariant two-form
$$
F_h \equiv \diff\left( \frac{h}{r^2} \right),
$$
enters the perturbation equations.

In terms of these
gauge-invariant quantities, the components of the linearized Einstein tensor
are
\bea
&& \delta G_{ab} = 0, \nonumber\\ && \delta G_{Ab}\,\diff x^b =
\left\{ \codtil\left[ r^4 \diff \left(\frac{h^{(inv)}}{r^2}\right) \right] +
\lambda h^{(inv)} \right\} \frac{S_A}{2r^2}, \label{Eq-DeltaGRW}\\
&& \delta
G_{AB} = - \codtil h^{(inv)} \,\nabhat_{(A} S_{B)}, \label{int_odd}
\eea
where the background equations have been used, and where here and in the following,
\bdm
\lambda \equiv (l-1)(l+2).
\edm

%%%%%%%%%%%%%%%%%%%%%%%%%%%%%%%%%%%%%%%%%%%%%%%%%%%%%%%%%%%%%%
\subsubsection{The master equation}
%%%%%%%%%%%%%%%%%%%%%%%%%%%%%%%%%%%%%%%%%%%%%%%%%%%%%%%%%%%%%%

The vacuum perturbations with odd parity are obtained from equation (\ref{Eq-DeltaGRW}),
which yields
\be
\codtil \left[ r^4 \diff \left( \frac{h^{(inv)}}{r^2} \right) \right] + \lambda h^{(inv)} = 0 .
\label{Eq-RW1}
\ee

The usual way to derive the RW equation for $l\geq 2$ from equation
(\ref{Eq-RW1}) is to decompose the one-form $h^{(inv)}$ with respect
to Schwarzschild coordinates, $h^{(inv)} = h^{(inv)}_t \diff t +
h^{(inv)}_r \diff r$, and to use the integrability condition
(\ref{int_odd}) to eliminate $h^{(inv)}_t$. This yields an equation
for $h^{(inv)}_r$ alone, which is then cast into a wave equation for
the function $\dot{\Phi}=(1-2M/r) h^{(inv)}_r/r$ ($\Phi$ defined
below).  This can also be achieved in a coordinate-invariant way as
follows: One uses the integrability condition $\codtil h^{(inv)} = 0$
to introduce the scalar potential $\Phi$ according to $h^{(inv)} =
\asttil\diff (r\Phi) = \epstil_{ab} \nabtil^a (r\Phi) \diff x^b$.
Equation (\ref{Eq-RW1}) may then be integrated to yield the following
wave equation

\be
\left[ -\laptil + r \laptil \left( \frac{1}{r}\right) +
\frac{\lambda}{r^2} \right] \Phi = 0,
\label{Eq-RW2}
\ee
where the two-dimensional Laplacian of a function is $\laptil \Phi
\equiv - \codtil \diff \Phi = \nabtil^a\nabtil_a \Phi$.  Here, the
free constant in the potential $\Phi$ has been used to set the
integration constant to zero.  Equation (\ref{Eq-RW2}) is the
coordinate-invariant version of the RW equation.  Indeed, we have not
specified any coordinates on the orbit manifold $\tilde{M}$.  Using
the coordinate-independent vacuum background equation $0 = r^2 G^a_{\;
a} = 2(r\laptil r + N - 1)$, equation (\ref{Eq-RW2}) finally assumes
the form

\be
\left[ -\laptil + V_{RW} \right] \Phi = 0,
\label{Eq-RW}
\ee
with
\bdm
V_{RW} = \frac{1}{r^2}\left[ l(l + 1) - \frac{6M}{r} \right].
\edm
>From equation (\ref{Eq-RW2}), we also get the following relation
\be
\Phi = -\frac{r^3}{\lambda} \asttil\diff\left( \frac{h^{(inv)}}{r^2} \right),
\label{Eq-RWPhi}
\ee
which enables us to compute $\Phi$ from the gauge-invariant one-form
$h^{(inv)}$.

For $l=1$ equation (\ref{Eq-RW2}) is immediately seen to admit the
solution $1/r$.  Since $\lambda = 0$, we may also directly integrate
equation (\ref{Eq-RW1}). This yields

\be
\asttil\diff\left( \frac{h}{r^2} \right) = -\frac{6J}{r^4},
\label{Eq-Kerr1}
\ee
where $6J$ is a constant of integration.  At this point, it is
important to recall that the one-form $h$ is {\it not\/}
coordinate-invariant, but transforms according to $h \mapsto h + r^2
\diff (f/r^2)$. This implies that the solution of the homogeneous part
of the above equation is pure gauge. A special solution is

\be
h = -\frac{2J}{r}\, \frac{\asttil\diff r}{N}\, .
\label{Eq-Kerr2}
\ee

As explicitly shown in \cite{SHB-Odd}, this describes the Kerr metric
in Boyer-Lindquist coordinates in first order of the rotation
parameter $a = J/M$. By equation (\ref{Eq-Kerr1}), $J$ is defined in a
coordinate-invariant way. In summary, a general $l=1$ perturbation is
given by

\be
h = -\frac{2J}{r}\, \frac{\asttil\diff r}{N} + r^2 \diff (f/r^2)
\ee
with $f$ an arbitrary function on the orbit space.

%%%%%%%%%%%%%%%%%%%%%%%%%%%%%%%%%%%%%%%%%%%%
\subsection{The even-parity sector}
%%%%%%%%%%%%%%%%%%%%%%%%%%%%%%%%%%%%%%%%%%%%
The even-parity perturbations of $g_{\mu\nu}$ are parameterized by a
symmetric tensor field $H_{ab}$, a one-form $Q_b$ and
two scalar fields $K$ and $G$ on the orbit space $\tilde{M}$,
\bea
\delta g_{ab} &=& H_{ab} Y, \nonumber\\
\delta g_{Ab} &=& Q_b \nabhat_A Y, \nonumber \\
\delta g_{AB} &=& K g_{AB} Y + G\, r^2\left( \nabhat_A\nabhat_B Y  +
\frac{1}{2}l(l+1)\ghat_{AB} Y \right). \nonumber
\eea
Here, the basis of symmetric tensors in $\delta g_{AB}$ is chosen to be
orthogonal with respect to the inner product induced by $\gtens$.
Furthermore, one has $\nabhat_A\nabhat_B Y + \frac{1}{2}l(l+1)\ghat_{AB} Y = 0$
for $l=0,1$; hence the amplitude $G$ is not present in those cases.
For $l=0$, the amplitude $Q_b$ is also absent.

%%%%%%%%%%%%%%%%%%%%%%%%%%%%%%%%%%%%%%%%%%%%
\subsubsection{Coordinate-invariant amplitudes}
%%%%%%%%%%%%%%%%%%%%%%%%%%%%%%%%%%%%%%%%%%%%
An infinitesimal coordinate transformation with even parity is
generated by a vector field $X$ with
$$
X^a = \xi^a Y, \;\;\;
X^A = f \ghat^{AB}\nabhat_B Y,
$$
where $\xi^a$ and $f$ are a vector field and a function, respectively,
on $\tilde{M}$. With respect to this, the metric perturbations transform
according to
\bea
H_{ab} &\mapsto& H_{ab} + \xi_{a|b} + \xi_{b|a}, \nonumber\\
   Q_b &\mapsto& Q_b + \xi_b + r^2 f_{|b}, \label{Eq-MCoordTransEven}\\
     K &\mapsto& K + 2 v^a \xi_a - l(l+1)f, \nonumber\\
     G &\mapsto& G + 2f. \nonumber
\eea
Here and in the following, $\xi_{b|a} \equiv \nabtil_a\xi_b$ denotes the
covariant derivative with respect to the orbit metric $\gtiltens$ and
$v_a \equiv r_{|a}/r\,$.

For $l\geq 2$, one can construct the following set of
coordinate-invariant amplitudes:
\begin{eqnarray}
H^{(inv)}_{ab} &=& H_{ab} - \left( p_{a|b} + p_{b|a} \right),
\label{Eq-CoordInvHKa} \\
K^{(inv)} &=& K - 2 v^a p_a +
\frac{1}{2}l(l+1) G, \label{Eq-CoordInvHKb}
\end{eqnarray}
where $p_a = Q_a - \frac{1}{2} r^2 G_{|a}$.
The (generalized) RW gauge is defined by choosing $\xi^a$ and
$f$ such that $Q_b$ and $G$ vanish. We see that in this gauge, which is also
unique, $H^{(inv)}_{ab}$ and $K^{(inv)}$ coincide with $H_{ab}$ and $K$.

For $l=1$, there is no such simple choice of coordinate-invariant
amplitudes, since $G$ is not present in this case.
Nevertheless, we can always chose the gauge such that $Q_b$ vanishes.
One then remains with $H_{ab}$ and $K$, which are subject to the
residual coordinate transformations as in
(\ref{Eq-MCoordTransEven}) with $\xi_b + r^2 f_{|b}
= 0$.

For $l=0$, $Q_b$ and $G$ are absent anyway and one can arrange
the gauge such that $K=0$.

In summary, it is sufficient to derive the linearized Einstein equations
for the perturbed metric
$$
\delta g_{ab} = H_{ab} Y, \;\;\;
\delta g_{Ab} = 0, \;\;\;
\delta g_{AB} = K g_{AB} Y,
$$
where for $l\geq 2$, $H_{ab}$ and $K$ can be replaced by their
coordinate-invariant counterparts defined in
(\ref{Eq-CoordInvHKa},\ref{Eq-CoordInvHKb}).

%%%%%%%%%%%%%%%%%%%%%%%%%%%%%%%%%%%%%%%%%%%%%%%%%%%
\subsubsection{The master equation}
%%%%%%%%%%%%%%%%%%%%%%%%%%%%%%%%%%%%%%%%%%%%%%%%%%%

The long but straightforward computation of the linearized Einstein
tensor is given in \cite{S-Diss}. The equations' structure
becomes much more transparent if one first splits the two-tensor
$H_{ab}$ into its trace and traceless part and then introduces
the one-form
\be
C = \hat{H}_{ab}\, r^{|a} \diff x^b,
\ee
where $\hat{H}_{ab}$ denotes the traceless part of $H_{ab}$.
A similar split is performed for the components of the Einstein tensor.
As a result, the relevant components of the Einstein tensor define
two scalars $S$ and $T$ and two one-forms $U$ and $V$ according to
\begin{eqnarray*}
& & \delta \hat{G}_{AB} = S \left( \nabhat_A\nabhat_B Y +
\frac{1}{2}l(l+1)\ghat_{AB} Y \right), \\
& & \gtil^{ab}\delta G_{ab} = T\, Y, \\
& & \delta G_{Ab}\,\diff x^b =
\frac{1}{2}\,U\,\nabhat_A Y, \\
& & \delta \hat{G}_{ab} r^{|a} \diff x^b = V\, Y.
\end{eqnarray*}

The vacuum field equations are then expressed in terms of the one-form
$C$ and the two scalars $H = \gtil^{ab} H_{ab}$ and $K$.  The simplest
equation, which is present only for $l\geq 2$, gives

\bdm
0 = -2S = H,
\edm
hence $H_{ab}$ is traceless (For $l=0,1$, we can make use of the
residual gauge freedom in order to impose $H=0$. Residual coordinate
transformations are then of the form (\ref{Eq-MCoordTransEven}) with
$\xi_a^{\; |a} = 0$ and $\xi_a = -r^2 f_{|a}$ for $l=1$.). Using
$H=0$, the remaining equations reduce to

\bea
0 = T &=& \frac{2}{r} \codtil C - \frac{2}{r^2} \gtiltens(C, \diff r) +
\laptil K + \frac{4}{r}\gtiltens(\diff K, \diff r)
        - \frac{\lambda}{r^2} K, \label{Eq-LEQ1}\\
0 = U &=& -\frac{1}{N} \left[ (\codtil C) \diff r + (\asttil\diff C) \asttil\diff r \right] -\diff K,
\label{EQ-LEQ2}\\
0 = V &=& (\codtil C) \frac{\diff r}{r} + \frac{1}{r} \diff\gtiltens( C, \diff r) +
\frac{l(l+1)}{2r^2} C  \nonumber\\
      &+& \frac{1}{2}\laptil K\,\diff r - \diff \gtiltens( \diff K, \diff r)
        + \left(\laptil r - \frac{N+1}{2r} \right)\diff K, \label{Eq-LEQ3}
\eea

where equation (\ref{EQ-LEQ2}) is void for $l = 0$. We recall that for
$l\geq 2$, we should replace $H_{ab}$ by $H^{(inv)}_{ab}$ and $K$ by
$K^{(inv)}$ in the above equations in order to give them a
coordinate-invariant meaning.  For $l\geq 1$, we compute the component
of $U$ parallel to $\asttil\diff r$:

\bdm
0 = \gtiltens(U, \asttil\diff r) = \asttil\diff C - \gtiltens(\diff K, \asttil\diff r)
  = \asttil\diff\left[ C - r\diff K \right].
\edm
This motivates us to replace the one-form $C$ with the one-form
\be
Z = C - r\diff K.
\label{Eq-ZerilliAmpl}
\ee
In terms of $Z$ and $K$ Einstein's equations become
\bea
&& 0 = \gtiltens(U, \asttil\diff r) = \asttil\diff Z, \label{Eq-LEQUT}\\
&& 0 = \gtiltens(U, \diff r) = -\codtil Z + r\laptil K, \label{Eq-LEQUR}\\
&& 0 = T = \frac{2}{r} \codtil Z - \frac{2}{r^2} \gtiltens(Z, \diff r) -
\laptil K - \frac{\lambda}{r^2} K, \label{Eq-LEQT}\\
&& 0 = r^2\left( 2V - T\diff r \right) = \diff\left[ 2r\gtiltens(Z, \diff r)
\right] + l(l+1) Z
 + r(a_0 + \lambda) \diff K + \lambda K \diff r, \label{Eq-LEQVT}
\eea
where we have defined
\bdm
a_0 = 2r\laptil r + 1 - N.
\edm
Using the background equation $0 = r^2 G^a_{\;\; a} = 2(r\laptil r + N - 1)$
and $N = 1 - 2M/r$, one finds $a_0 = 6M/r$.

In view of equation (\ref{Eq-LEQUT}), we may introduce the scalar
field $\zeta$ according to $Z = \diff\zeta$. Equation (\ref{Eq-LEQVT})
may then be integrated to yield

\be
2r\gtiltens(\diff\zeta, \diff r) + l(l+1) \zeta + r(a_0 + \lambda) K = 0.
\label{Eq-LEQVTbis}
\ee

It is now clear how to obtain a single, second order differential
equation for $\zeta$: First, we eliminate $\laptil K$ from equations
(\ref{Eq-LEQUR}) and (\ref{Eq-LEQT}). This gives

\be
-\laptil\zeta - \frac{2}{r} \gtiltens(\diff\zeta, \diff r) - \frac{\lambda}{r} K = 0.
\label{Eq-ElimZerilli}
\ee

Next, this equation is used to eliminate $K$ in
(\ref{Eq-LEQVTbis}). Hence,

\be
-(a_0 + \lambda) \laptil\zeta - \frac{2a_0}{r} \gtiltens(\diff\zeta, \diff r) +
\frac{l(l+1)\lambda}{r^2} \zeta = 0.
\label{Eq-PreZerilli}
\ee
(Note that for $l=1$, this equation is equivalent to equation
(\ref{Eq-ElimZerilli}) and thus is also valid in that case.)  Finally,
we define the new scalar function $\Psi$ by

\bdm
\zeta = (a_0 + \lambda)\Psi,
\edm
in order to remove the first order derivatives. This yields the
Zerilli equation \cite{Zerilli},

\be
\left[ -\laptil + V_Z \right] \Psi = 0,
\label{Eq-Zerilli}
\ee
where

\be
V_Z = \frac{ \lambda^2 r^2 [ (\lambda+2)r + 6M) ] + 36 M^2 (\lambda r + 2M) }{(\lambda r + 6M)^2 r^3}\, .
\label{Eq-ZerilliPotential}
\ee

Before we discuss the special cases $l=0$ and $l=1$, we make two
remarks: First, for $l\geq 2$, the scalar field $\Psi$ can be obtained
from the Zerilli one-form $Z$ using equation (\ref{Eq-PreZerilli}).
The second point is that it is also possible to obtain the RW equation
for the scalar $\Phi_e = r^2\codtil Z$. In fact, Chandrasekhar (see,
e.g. \cite{C-Book}) has shown that the equations of RW and Zerilli for
a Schwarzschild background are equivalent in the frequency domain.
However, in the time domain, we were not able to express $Z$ in terms
of $\Phi_e$ and its derivatives alone. For this reason, we will use
the Zerilli equation in the even-parity sector and not the RW
equation.

For $l=1$, equation (\ref{Eq-PreZerilli}) reduces to
\be
-\laptil\zeta - \frac{2}{r} \gtiltens(\diff\zeta, \diff r) = \frac{1}{r^2} \codtil( r^2 \diff\zeta) = 0.
\label{Eq-PreZerillil1}
\ee

However we recall that for $l=1$, $H_{ab}\,$, and hence also $\zeta$
are not defined in a coordinate-invariant manner. Under the residual
gauge-freedom (\ref{Eq-MCoordTransEven}) with $\xi_a = -r^2 f_{|a}$
and

\be
\codtil( r^2\diff f) = -\xi_a^{\; |a} = 0,
\label{Eq-ResGaugeCond}
\ee
we find that $Z$ transforms according to
\bdm
Z \mapsto Z + 6M\,\diff f, \quad\hbox{and hence} \quad
\zeta\mapsto\zeta + 6M f.
\edm

Since $f$ is an arbitrary solution of equation
(\ref{Eq-ResGaugeCond}), and since the equations
(\ref{Eq-PreZerillil1}) and (\ref{Eq-ResGaugeCond}) are equivalent, it
is clear that every solution of (\ref{Eq-PreZerillil1}) corresponds to
pure gauge. In particular, we can choose the residual gauge in order
for $\zeta$ to vanish. In this gauge $K$ vanishes as well, as a
consequence of equation (\ref{Eq-LEQVTbis}). The even-parity sector is
therefore empty for $l=1$.

For $l=0$, one can choose the gauge such that both $H$ and $K$ vanish.
Then equations (\ref{Eq-LEQ1}) and (\ref{Eq-LEQ3}) yield
\bdm
\codtil (rC) = 0, \;\;\;
\gtiltens(rC, \diff r) \equiv 2\delta M = \const,
\edm
which has the general solution
\be
C = \frac{2\delta M}{rN}\diff r + \asttil\diff h.
\label{Eq-DMC}
\ee
Here, $\delta M$ is a constant describing the variation of the ADM
mass, and $h$ is a function that only depends on $r$.  Comparing this
with a residual gauge, which is generated by $\xi_a = \epstil_{ab}
k^{|b}$ for a function $k$ of $r$, we get

\bdm
C \mapsto C - \asttil( N k''(r) \diff r),
\edm
showing that the function $h(r)$ above corresponds to pure gauge.
This can also be seen in a gauge-invariant way: Recall that for any
spherically symmetric metric of the form (\ref{Eq-Metric}) we defined
the mass parameter $M$ through $1 - 2M/r = N = \gtil^{ab} r_{|a}
r_{|b}\,$.  Using the fact that (for Y=1) $\delta\gtil_{ab} = H_{ab}$
and $\delta(r^2) = r^2 K$, we obtain
\be
2\delta M = r r^{|a} r^{|b} H_{ab} - r r^{|b} (rK)_{|b} + M K.
\label{Eq-DMGaugeInv}
\ee
It can be checked that the RHS is indeed a gauge-invariant
combination. On the other hand, for $K = 0$, equation
(\ref{Eq-DMGaugeInv}) yields $2\delta M = \gtiltens(rC, \diff r)$,
as above.

%%%%%%%%%%%%%%%%%%%%%%%%%%%%%%%%%%%%%%%%%%%%%%%%%%%%
\subsection{Summary}
%%%%%%%%%%%%%%%%%%%%%%%%%%%%%%%%%%%%%%%%%%%%%%%%%%%%

In both the odd- and the even-parity sector, perturbations on any
spherically symmetric vacuum background are described by a wave
equation of the form
\be
\left( -\laptil + V \right) u = 0,
\label{master_lapla}
\ee
where $\laptil$ is the Laplacian with respect to the orbit metric
$\gtiltens$ and where the potential $V$ depends on the ADM mass $M$,
$r$ and the angular momentum number $l$ only.

For $l=0$ and $l=1$, there are no dynamical modes. The only physical
solutions in those cases are stationary, describing variation of the
mass and angular momentum.  The gauge-invariant part of the metric can
be obtained from $u$ and vice-versa. These relations are going to be
made more precise in the next section.

Finally, we would like to mention that our gauge-invariant
perturbation formalism has also been generalized to the case where
matter fields are coupled to the metric \cite{S-Diss,SHB-Odd}. In
the case of Einstein-Maxwell, we were able to generalize the
equations obtained by Moncrief \cite{Moncrief}. However, as we
have argued in a recent Letter \cite{BHS-Letter}, the perturbation
formalism presented here fails to yield a wave equation of the
form (\ref{master_lapla}) with a {\it symmetric\/} potential,
$V = V^T$, when non-Abelian fields are coupled to the metric.

%%%%%%%%%%%%%%%%%%%%%%%%%%%%%%%%%%%%%%%%
\section{Relation to the ADM quantities}
%%%%%%%%%%%%%%%%%%%%%%%%%%%%%%%%%%%%%%%%

As mentioned in the introduction, one of the motivations for the
present work is Cauchy-perturbative matching. This amounts to
matching numerically, at each time step, the variables used in a
nonlinear code with the ones used in the perturbative regime (in our
case the RW and Zerilli functions). The matching takes place at a
timelike boundary. For this purpose we explicitly show the relation
between the RW and Zerilli gauge-invariant potentials and the ADM
quantities, namely, the three-metric and the extrinsic curvature. This
does not restrict the formulation of Einstein's equations to be
used in the nonlinear regime, since for a formulation other than the
standard ADM (e.g. conformal ADM, or a hyperbolic formulation) the
relevant quantities can be obtained from the three-metric and the
extrinsic curvature, and vice-versa.

So our aim is to make explicit the relationship between the scalar
fields $\Phi$ and $\Psi$ satisfying the RW and Zerilli equations
(\ref{Eq-RW}) and (\ref{Eq-Zerilli}) and the components of the
linearized $3$-metric and extrinsic curvature.  We will show in this
section that - modulo gauge transformations - there is a one-to-one
correspondence between $\delta\gbar_{ij}$, $\delta K_{ij}$ and the
scalar amplitudes $\Phi$, $\dot{\Phi} \equiv \p_t\Phi$, $\Psi$,
$\dot{\Psi}$.  Furthermore, this correspondence involves no
time-derivatives.  For example, it is possible to express $\dot{\Psi}$
in terms of purely spatial quantities, i.e. $\delta\gbar_{ij}$,
$\delta K_{ij}$ and their {\it spatial} derivatives only.

We assume that the full metric, satisfying the nonlinear field
equations, has the ADM form

\bdm
\gtens(\mu) = -\alpha(\mu)^2 \diff t^{\, 2} +
\gbar_{ij}(\mu) \left( \diff x^i + \beta^i(\mu) \diff t \right)
\left( \diff x^j + \beta^j(\mu) \diff t \right),
\edm
where $\mu$ is a variational parameter, such that for $\mu=0$, the
metric is spherically symmetric. With respect to the 2+2 split
(\ref{Eq-Metric}), the orbit metric $\gtiltens$ takes the form

\bdm
\gtiltens = -\alpha^2\, \diff t^{\, 2}
 + \gamma^2\left( \diff x + \beta \diff t \right)^2,
\edm
where $x$ is any radial coordinate, $\alpha$ and $\beta \equiv
\beta^x$ are the background lapse and shift, respectively, and
$\gamma^2 \equiv \gbar_{xx}$.  The components of the extrinsic
curvature are

\begin{eqnarray*}
2\alpha K_{xx} &=& 2\gamma\left(\p_0\gamma - \gamma\beta' \right), \\
2\alpha K_{xA} &=& 0, \\
2\alpha K_{AB} &=& 2r\p_0 r\,\ghat_{AB},
\end{eqnarray*}
where a prime denotes differentiation with respect to $x$, and where
we have also introduced the normal derivative $\p_0 \equiv \p_t -
\beta\p_x$.

The components of the linearized metric have the form

\bea
\delta g_{tt} &=& -2\alpha\delta \alpha - \beta^2 \delta\gbar_{xx} +
2\beta \delta\beta_x\, , \nonumber\\
\delta g_{tj} &=& \delta\beta_j, \label{Eq-DgDef}\\
\delta g_{ij} &=& \delta\gbar_{ij}, \nonumber
\eea
where $i=x,A$.  Note that we use perturbations of the coshift rather
than the shift vector.  This fact will turn out to be important when
we try to express the ADM quantities in terms of the RW and Zerilli
scalars.  Similarly, the components of the linearized extrinsic
curvature are given by

\bea
2\alpha\delta K_{xx} &=& \p_t \delta\gbar_{xx} - 2K_{xx}\delta\alpha
+ \beta\gamma^2\left( \frac{\delta\gbar_{xx}}{\gamma^2} \right)' - 2\gamma\left( \frac{\delta\beta_x}{\gamma}\right)',
\nonumber\\
2\alpha\delta K_{xA} &=& \p_t \delta\gbar_{xA} -
2\beta\frac{r'}{r}\delta\gbar_{xA} + \beta\nabhat_A \delta\gbar_{xx} -
\nabhat_A \delta\beta_x - r^2\left( \frac{\delta\beta_A}{r^2} \right)',
\nonumber\\
2\alpha\delta K_{AB} &=& \p_0 \delta\gbar_{AB} - 2K_{AB}\delta\alpha
+ 2\beta \nabhat_{(A} \delta g_{B) x} - 2\nabhat_{(A}\delta\beta_{B)}\label{Eq-DKDef}\\
&-& 2\frac{r r'}{\gamma^2} \ghat_{AB} \left( \delta\beta_x - \beta\delta\gbar_{xx} \right).
\nonumber
\eea

%%%%%%%%%%%%%%%%%%%%%%%%%%%%%%%%%%%%%%
\subsection{The odd-parity sector}
%%%%%%%%%%%%%%%%%%%%%%%%%%%%%%%%%%%%%%

In the odd-parity sector with $l\geq 2$, the only non-vanishing
perturbations can be parameterized according to

\begin{eqnarray*}
&& \delta\beta_A = b S_A, \\
&& \delta\gbar_{xA} = h_1 S_A, \;\;\;
   \delta\gbar_{AB} = 2h_2 \nabhat_{(A} S_{B)}, \\
&& \delta K_{xA} = \pi_1 S_A, \;\;\;
   \delta K_{AB} = 2\pi_2 \nabhat_{(A} S_{B)}.
\end{eqnarray*}

%%%%%%%%%%%%%%%%%%%%%%%
\subsubsection{The potentials in term of the ADM quantities ($l \geq 2$)}
%%%%%%%%%%%%%%%%%%%%%%%

We want to express $\Phi$ and $\dot{\Phi}$ in terms of the quantities
$b,h_1,h_2,\pi _1,\pi _2$ and their spatial derivatives.

First, we observe that $h_t = b$, $h_x = h_1$, and $k = h_2$ where
$h_t$, $h_x$ and $k$ are the amplitudes introduced in
(\ref{Eq-OddDeltag}). Therefore, we obtain

\bea
h^{(inv)}_t &=& b - \dot{h_2} + 2\frac{\dot{r}}{r} h_2 \label{hdert} \\
h^{(inv)}_x &=& h_1 - h_2' + 2\frac{r'}{r} h_2\, . \label{hderx}
\eea

Next, the equations (\ref{Eq-DKDef}) yield the relations

\begin{eqnarray}
2\alpha\,\pi_1 &=& \dot{h}_1 - 2\beta\frac{r'}{r} h_1 -
r^2\left( \frac{b}{r^2} \right)', \label{pi1} \\
2\alpha\,\pi_2 &=& \p_0 h_2 + \beta h_1 - b\, .\label{pi2}
\end{eqnarray}

Using the last of these relations, eq. (\ref{pi2}), to re-express, in
eq. (\ref{hdert}), time derivatives of $h_2$ in terms of spatial
quantities, the components of the gauge-invariant one-form $h^{(inv)}$
take the form

\bea
 & & h^{(inv)}_0 = - 2\alpha\,\pi_2 + 2\frac{\partial _0r}{r} h_2,
 \label{hinv0}\\
& & h^{(inv)}_x = h_1 - r^2\left( \frac{h_2}{r^2} \right)',
\label{hinvx}
\eea
where $h^{(inv)}_0 \equiv h^{(inv)}_t - \beta h^{(inv)}_x$.  Next, one
uses equation (\ref{Eq-RWPhi}), trading time derivatives for spatial
ones with the aid of (\ref{pi1}), to find

\be
\Phi = \frac{r}{\lambda\alpha\gamma} \left( 2\alpha\,\pi_1 -  2\frac{\p_0
r}{r} h_1 \right), \label{pot_adm}
\ee
which is one of the formulae we were looking for.  In order to obtain
the time derivative of $\Phi$, one uses the definition of $\Phi$,
i.e. $h^{(inv)} = \asttil\diff( r\Phi )$.  This yields

\begin{eqnarray}
h^{(inv)}_0 &=& -\frac{\alpha}{\gamma} \p_x (r\Phi), \label{Eq-HPhia}\\
h^{(inv)}_x &=& -\frac{\gamma}{\alpha} \p_0 (r\Phi).
\label{Eq-HPhib}
\end{eqnarray}
which one can solve for $\dot{\Phi }$. Using
eqs.(\ref{pot_adm},\ref{hinv0},\ref{hinvx}), the result is

\be
\dot{\Phi} = \frac{1}{\gamma r}\left( -\alpha + \frac{2\dot{r}\partial
_0r}{\lambda \alpha } \right)h_1  + \frac{2\gamma \beta}{r}\pi _2 -
\frac{2\dot{r}}{\lambda \gamma } \pi _1 + \frac{r\alpha }{\gamma
}\left(\frac{h_2}{r^2} \right)^{'} - \frac{2\gamma \beta \dot{r}}{r^2\alpha
}h_2
\ee

%%%%%%%%%%%%%%%%%
\subsubsection{The ADM quantities from the potentials ($l \geq 2$)}
%%%%%%%%%%%%%%%%%

On the other hand, given $\Phi$ and $\dot{\Phi}$, we obtain $\delta
g_{ij}$ and $\delta K_{ij}$ in the following way: First, we compute
$h^{(inv)}_0$ and $h^{(inv)}_x$ from equation
(\ref{Eq-HPhia},\ref{Eq-HPhib}). Then, using the above equations, it
is straightforward to express $b$, $h_1$, $\pi_1$ and $\pi_2$ in terms
of $h^{(inv)}_0$, $h^{(inv)}_x$, $\Phi$ and $k$, where $k$
parameterizes the gauge freedom:

\begin{eqnarray}
& & b = h^{(inv)}_0 + \beta h^{(inv)}_x + r^2\p_t\left( \frac{k}{r^2}
\right), \label{k_shift_odd}\\
& & h_1 = h^{(inv)}_x + r^2\p_x\left( \frac{k}{r^2}\right), \\
& & h_2 = k, \\
& & 2\alpha\pi_1 = \lambda\frac{\alpha\gamma}{r}\Phi + 2\frac{\p_0 r}{r}
h_1\, , \\
& & 2\alpha\pi_2 = -h^{(inv)}_0 + 2\frac{\p_0 r}{r} k.
\end{eqnarray}

%%%%%%%%%%%
\subsubsection{The special case $l =1$}
%%%%%%%%%%%

For $l=1$, the amplitudes $h_2$ and $\pi_2$ are absent.  According to
the analysis in the last section, the only physical solution is the
Kerr mode.  Using equation (\ref{Eq-Kerr1}), one finds that the
rotation parameter (the only gauge invariant for $l =1$ ) can be
extracted from the ADM quantities according to

\be
6J = \frac{r^2}{\alpha\gamma} \left( 2\alpha\,\pi_1 - 2\frac{\p_0 r}{r} h_1
\right). \label{Eq-Geta}
\ee

On the other hand, using (\ref{Eq-Kerr2}), one finds

\bea
& &
b = \frac{2J}{rN} \left( \frac{\beta\gamma}{\alpha} \p_0 r +
\frac{\alpha}{\gamma} r' \right) + r^2\p_t \left( \frac{f}{r^2} \right),
\label{k_shift_1}\\
& & h_1 = \frac{2J}{rN} \frac{\gamma}{\alpha} \p_0 r + r^2\p_r \left(
\frac{f}{r^2} \right), \nonumber
\eea
where $f$ parameterizes the gauge freedom, and where $N = 1 - 2M/r =
-(\p_0 r)^2/\alpha^2 + r'^2/\gamma^2$.  The amplitude $\pi_1$ then
follows from (\ref{Eq-Geta}).

%%%%%%%%%%%%%%%
\subsection{The even-parity sector}
%%%%%%%%%%%%%%%%%

Here the perturbations are

\begin{eqnarray*}
&& \delta\alpha = a\, Y, \;\;\;
   \delta\beta_x = b_1 Y, \;\;\;
   \delta\beta_A = b_2 \nabhat_A Y,\nonumber\\
&& \delta\gbar_{xx} = h\, Y, \;\;\;
   \delta\gbar_{xA} = q \nabhat_A Y, \nonumber\\
&& \delta\gbar_{AB} = K \gbar_{AB} Y + G\, r^2\left( \nabhat_A\nabhat_B Y +
\frac{1}{2}l(l+1)\ghat_{AB} Y \right),\\
&& \delta K_{xx} = \pi_h\, Y, \;\;\;
   \delta K_{xA} = \pi_q \nabhat_A Y, \nonumber\\
&& \delta K_{AB} = \pi_K \gbar_{AB} Y + \pi_G\, r^2\left( \nabhat_A\nabhat_B Y +
\frac{1}{2}l(l+1)\ghat_{AB} Y \right).
\end{eqnarray*}

%%%%%%%%%%%%%%%%%%%%%%%%%%%%%%%%%%%%%%
\subsubsection{The potential in terms of the ADM quantities ($l \geq 2$)}
%%%%%%%%%%%%%%%%%%%%%%%%%%%%%%%%%%%%%

>From equations (\ref{Eq-DgDef}) one finds $H_{tt} = -2\alpha\, a - \beta^2 h + 2\beta b_1$,
$H_{tx} = b_1$, $H_{xx} = h$, $Q_t = b_2$, $Q_x = q$, while $K$ and $G$ agree with their
definitions in the previous section.  The expressions for the linearized curvature tensor,
equation (\ref{Eq-DKDef}), yield

\bea
2\alpha\,\pi_h &=& \dot{h} + \beta\gamma^2 \left( \frac{h}{\gamma^2} \right)'
- 2\gamma\left( \frac{b_1}{\gamma} \right)' - \frac{2a\gamma }{\alpha
}(\partial _0-\gamma \beta '), \nonumber\\
2\alpha\,\pi_q &=& \dot{q} - 2\beta\frac{r'}{r} q +\beta h - b_1 - r^2 \left( \frac{b_2}{r^2} \right)',
\label{Eq-DPiEven}\\
2\alpha\,\pi_K &=& \frac{1}{r^2}\p_0 (r^2 K) + \frac{2r'}{r\gamma^2} \left( \beta h - b_1 \right)
- \frac{l(l+1)}{r^2} \left( \beta q - b_2 \right) - \frac{2}{\alpha r} \p_0 r\, a,
\nonumber\\
2\alpha\,\pi_G &=& \frac{1}{r^2}\p_0 (r^2 G) + \frac{2}{r^2}\left( \beta q - b_2 \right).
\nonumber
\eea

At first sight it is not clear how the Zerilli one-form $Z$, defined
in (\ref{Eq-ZerilliAmpl}) can be expressed in terms of spatial
amplitudes only, since from the definition of $H^{(inv)}_{ab}$ one
sees that second time derivatives of metric components can appear.
However, it turns out that only the two-form $\omega_{ab} = p_{b|a} -
p_{a|b}$, which contains no second time derivatives of $h$, $q$, $K$
and $G$, appears in the Zerilli one-form.  Using the fact that
$H^{(inv)}_{ab}$ is traceless as a consequence of the field equations,
one obtains

\bdm
Z_a = H_{ab} r^{|b} - r K_{|a} - \frac{1}{2}l(l+1) r G_{|a}
+ r^{|b}\omega_{ab} + 2r v_{b|a} p^b.
\edm
Now, using (\ref{Eq-DPiEven}), it is easy to find
\bea
p_0 &=& -\alpha r^2\pi_G + r(\p_0 r) G, \label{eq-p0}\\
p_x &=& q - \frac{1}{2} r^2 G', \label{eq-px}\\
\omega_{0x} &=& 2\alpha( \pi_q + rr'\pi_G ) -\frac{1}{r} (\p_0 r) (r^2 G)'
- (\beta h - b_1). \nonumber
\eea

Using the background equation $v_{b|a} = M r^{-3}\gtil_{ab} - r^{-2}
r_{|a}r_{|b}$, one eventually obtains

\bea
Z_0 &=& -2\alpha r\left( \pi_K + \frac{1}{2}l(l+1)\pi_G \right)
     + 2(\p_0 r) \left( K + \frac{1}{2}l(l+1)G\right)
\label{z0}\\
    &+& \frac{r'}{\gamma^2} \left( 2\alpha \pi_q - 2\frac{\p_0 r}{r} q  \right)
 - \frac{2}{r} \left( 1 -3 \frac{M}{r} \right) p_0\,
, \nonumber\\
Z_x &=& \frac{r'}{\gamma^2} h - r\left( K + \frac{1}{2}l(l+1)G\right)'
+ \frac{\p_0 r}{\alpha^2} \left( 2\alpha \pi_q - 2\frac{\p_0 r}{r} q \right)
  - \frac{2}{r} \left( 1 - 3\frac{M}{r} \right) p_x\,
. \label{zx}
\eea

The Zerilli scalar $\zeta$ (and $\Psi$) can now be obtained from its
definition, eq.(\ref{Eq-LEQVTbis}), with

$$
K^{(inv)} = K + \frac{1}{2}l(l+1)G - \frac{2}{r} r^{|b} p_b\, ,
$$
$$
r^{|b} p_b = -(\p_0 r) p_0/\alpha^2 + r' p_x/\gamma^2 \, ,
$$
and the one-form $Z$ given by equations (\ref{z0},\ref{zx}). On the
other hand, the latter equations also give us $\dot{\zeta}$ from
$\dot{\zeta} = Z_0 + \beta Z_x$. Note that -- as in the odd-parity
sector -- the scalars $\zeta$ and $\dot{\zeta}$ do not depend on the
perturbed lapse nor on the perturbed shift.  Finally, we see that for
a Schwarzschild slicing where $\p_0 r = 0$, $\Phi$ and $\dot{\zeta}$
are linear combinations of the extrinsic curvature components only. These
combinations precisely agree with the ones obtained in a perturbative
approach on a static background in terms of curvature-based quantities
\cite{Curv-Pert}.

%%%%%%%%%%%%%%%%%%%%%%%%%%%%%%%%
\subsubsection{The ADM quantities from the potentials ($l \geq 2$)}
%%%%%%%%%%%%%%%%%%%%%%%%%%%%%%%%

If $\zeta$ and $\dot{\zeta}$ are known, equation (\ref{Eq-LEQVTbis})
tells us how to obtain $K^{(inv)}$ and $\dot{K}^{(inv)}$. Next, the
traceless part of $H^{(inv)}_{ab}$ is obtained from this and the
definition of the Zerilli one-form $Z$.  Finally, one has

\bea
& & 2\alpha\, a = -H^{(inv)}_{00} - 2p_{t|t} + 2\beta( p_{t|x} + p_{x|t} ) -
2\beta^2 p_{x|x},\label{a_ptt}\\
& & h = H^{(inv)}_{xx} + 2p_{x|x}, \nonumber\\
& & b_1 = H^{(inv)}_{tx} + p_{t|x} + p_{x|t}, \label{b1_pxt}\\
& & b_2 = p_t + \frac{1}{2} r^2 \dot{G}, \label{b2_gt}\\
& & q = p_x + \frac{1}{2} r^2 G', \nonumber\\
& & K = K^{(inv)} + \frac{2}{r} r_{|b} p^b - \frac{1}{2}l(l+1) G.
\nonumber
\eea
Here, $p_a$ and $G$ parameterize the gauge freedom.
The amplitudes $\pi_h$, ... $\pi_G$ are obtained from this and
(\ref{Eq-DPiEven}).

%%%%%%%%%%%%%%%%%%%%%%%%%%%%%%%%
\subsubsection{The special case $l =1$}
%%%%%%%%%%%%%%%%%%%%%%%%%%%%%%%%

For $l=1$, according to the analysis of the last section, we have only
gauge modes, and $G$ and $\pi_G$ are absent.  Therefore, the ADM-based
amplitudes are obtained from the same equations as above, but where
$H^{(inv)}_{ab}$ and $K^{(inv)}$ are set to zero, and $p_a$ and $G$
are replaced by $\xi_a$ and $2f$, respectively, where $\xi^a$ and $f$
parameterize the gauge transformation that brings us from the RW
gauge to the actual gauge that one wants to use.

%%%%%%%%%%%%%%%%%%%%%%%%%%%%%%%%
\subsubsection{The special case $l =0$}
%%%%%%%%%%%%%%%%%%%%%%%%%%%%%%%%

For $l=0$, $b_2$, $q$ and $\pi_q$ are also absent.  Using equation
(\ref{Eq-DMGaugeInv}) and the relations (\ref{Eq-DPiEven}), the
perturbed mass parameter is found to be

\bdm
\delta M = \frac{r^2}{\alpha} (\p_0 r) \pi_K
 + \frac{r}{2}\left( \frac{r'}{\gamma^2} \right)^2 h
 + \frac{1}{2} (r - M) K - \frac{r'}{2\gamma^2} (r^2 K)'.
\edm

In order to obtain the perturbed three-metric and extrinsic curvature
in terms of $\delta M$, one uses equation (\ref{Eq-DMC}) which gives

\bdm
H_{ab} = \frac{4\delta M}{r N^2} \left( r_{|a} r_{|b} - \frac{N}{2} \gtil_{ab} \right)
+ \hbox{gauge},
\edm
and the ADM quantities are obtained in a similar way to above.

%%%%%%%%%%%%%%%%%%%%%%%%%%%%%%%%
\subsection{Gauge fixing vs choices of lapse and shift}
%%%%%%%%%%%%%%%%%%%%%%%%%%%%%%%%

We have shown above how to construct the three metric and extrinsic
curvature from the potentials (and vice-versa), up to gauge
freedom. In numerical simulations, however, usually one does not fix
the gauge but rather chooses lapse and shift, perhaps as prescribed
functions of spacetime (``exact lapse'' or ``exact shift'') or as
dynamical quantities coupled to the the three metric and/or extrinsic
curvature (``live gauges''). In general this {\em does not fix the
gauge completely}, which means that we have to relate the gauge
freedom to the choice of lapse and shift.  The properties of such
relations depend on the details of how the lapse and shift are
chosen, and it is therefore not possible to give a general
discussion. These equations, for example, might be elliptic if some
kind of minimal distortion is imposed, hyperbolic as in the case we
discuss below, or of some other (perhaps unknown) type.

Here we will concentrate on a specific simple prescription, but it
should be clear that other cases can be treated similarly. The case we
are going to discuss is exact-coshift, exact-lapse; that is, the lapse
and shift covector are arbitrary but {\it a priori} given functions on the
orbit space.

We start with the odd-parity sector. The perturbed lapse is zero, and,
for $l \geq 2$, the perturbed coshift is given by the right hand side
of equation (\ref{k_shift_odd}). The function $k$ parameterizes the
gauge freedom, and it is thus given by the equation

\be
\partial _t k = \frac{1}{r^2}\left(b-h_0^{inv} - \beta h_x^{inv}\right)
\label{time_k}
\ee
Since $b$ is a given function, this equation can be solved for $k$,
provided we supply initial data. Given any three-metric and
extrinsic curvature at $t_0$, the initial data for $k$ is given by $k(t_0,x)=h_2(t_0,x)$.

The treatment for $l =1$ is similar. Now the gauge freedom is
controlled by $f$, which can be related to the coshift by equation
(\ref{k_shift_1}), rewritten as

\be \partial
_t\left(\frac{f}{r^2} \right) = \frac{1}{r^2}\left[ b - \frac{2J}{rN} \left(
\frac{\beta\gamma}{\alpha} \p_0 r + \frac{\alpha}{\gamma} r' \right) \right] \label{time_k_1} \ee

In the even-parity case with $l \geq 2$ the gauge functions $p_t,p_x$
and $G$ are related to the lapse and shift by evolution equations
which are straightforwardly obtained from equations
(\ref{a_ptt},\ref{b1_pxt},\ref{b2_gt}). These evolution equations form
a $3 \times 3$ coupled system, first order in space and in time,

\begin{eqnarray}
\dot{G} &= & 0 + \mbox{l.o.} \label{gdot}\\
\dot{p_x} &= & -p_t' + \mbox{l.o.} \label{pxdot}\\
\dot{p_t} &= & -\beta ^2 p_x' + \mbox{l.o.} \label{ptdot}
\end{eqnarray}
where l.o. stands for lower order terms.
Initial data for the system (\ref{gdot},\ref{pxdot},\ref{ptdot}) is
given by the three-metric and the extrinsic curvature at some time $t_0$
and the formulas (\ref{eq-p0},\ref{eq-px}) for $p_t$ and $p_x$.
It is easy to see that
equations (\ref{gdot},\ref{pxdot},\ref{ptdot}) constitute a weakly
hyperbolic (see, eg, \cite{kreiss_lorentz}) system if $\beta =0$ and a
stricly hyperbolic system otherwise. That is, if these equations are
written as $u_t = Au_x +$ l.o., with $u = (G,p_x,p_t)^T$, the matrix
$A$ has three different real eigenvalues if $\beta\neq 0$; and a
single degenerate real eigenvalue (zero) with only two independent
eigenvectors if $\beta =0$. The structure of the equations for $l =1$
is the same, replacing $p_a,G$ by $\xi_a, 2f$, respectively.  Finally,
for $l =0$ the situation is similar but simpler: $G$ does not appear,
and the principal part of the evolution equations for two gauge
quantities $p_x$ and $p_t$ is also given by
$(\ref{pxdot},\ref{ptdot})$. As before, these equations are weakly
hyperbolic if the background shift is zero, and strictly hyperbolic
otherwise.  If we use densitized lapse, as is usually done in
hyperbolic formulations (see, e.g., \cite{Hyperbolic}), the above
system of equations is strongly hyperbolic even in the case where
$\beta = 0$. In contrast to this, the system is ill posed if we use
exact shift instead of coshift. For $l = 0$, this fact has already
been noted in \cite{1D-paper}.

%%%%%%%%%%%%%%%%%%
\section{Relation to the Teukolsky formalism}
%%%%%%%%%%%%%%%%%%%
\label{Sect-4}

In order to compare our perturbation equations with the Teukolsky
equation for a non-rotating background, we introduce a NP null tetrad
that is adapted to the spherically symmetric metric (\ref{Eq-Metric}),
i.e.

\bdm
l = l_a\diff x^a, \;\;\;
n = n_a\diff x^a, \;\;\;
m = m_A\diff x^A,
\edm
where $l$ and $n$ form a null dyad of $\gtiltens$,

\bdm
\gtil_{ab} = -l_a n_b - l_b n_a\, ,
\edm
and $m$ is a complex one-form such that

\bdm
r^2\ghat_{AB} = m_A\bar{m}_B +  m_B\bar{m}_A\, .
\edm

Here and in the following, a bar denotes complex conjugation.  Note
that

\bdm
\epstil_{ab} = l_a n_b - l_b n_a\, , \;\;\;
r^2\epshat_{AB} = i\left( m_A\bar{m}_B -  m_B\bar{m}_A \right).
\edm

The only non-vanishing NP coefficients are

\bea
&& \rho = -\frac{1}{r} D r, \;\;\;
   \mu = \frac{1}{r} \Delta r, \nonumber\\
&& \epsilon = \frac{1}{2} \diff l(l,n) = \frac{1}{2} l^a n^b\nabtil_a l_b\, ,
\nonumber\\
&& \gamma = \frac{1}{2} \diff n(l,n) = -\frac{1}{2} l^b n^a\nabtil_a n_b\, , \nonumber\\
&& \alpha = -\bar{\beta} = \frac{1}{r} \hat{\alpha}, \nonumber
\eea
where $\hat{\alpha} = -\frac{1}{2} \diff\bar{\hat{m}}
(\hat{m},\bar{\hat{m}}) = \frac{1}{2}
\bar{\hat{m}}^A\hat{m}^B\nabhat_A\bar{\hat{m}}_B$ is a NP coefficient
with respect to the dyad defined by $\hat{m}\equiv\frac{1}{r} m$.
Here, $D = l^a\nabtil_a$ and $\Delta = n^a\nabtil_a\,$.  We also
introduce, for later use, the angular derivative operator
$\hat{\delta} = \hat{m}^A \nabhat_A\,$. From the NP vacuum equations
(see, e.g. \cite{S-NP}), it then follows that all Weyl scalars but
$\Psi_2$ vanish, $\Psi_0 = \Psi_1 = \Psi_3 = \Psi_4 = 0$.  In terms of
the invariant definition of $M$ given in section 2, $\Psi_2$ can be
expressed as

\bdm
\Psi_2 = -\frac{M}{r^3}\, .
\edm

In particular, the metric is of type D with repeated principal null
vectors aligned with $l^a$ and $n^a$.

In what follows, we study the decoupled equation derived by Teukolsky
\cite{Teukolsky} governing linear fluctuations of $\Psi_4$ on any
spherically symmetric vacuum background. To linear order, $\Psi_0$ and
$\Psi_4$ are invariant with respect to both infinitesimal coordinate
transformations and null tetrad rotations. The reason why we focus on
$\Psi_4$ and not $\Psi_0$ is that we want to study outgoing radiation
at null infinity, which is described by $\Psi_4$ (see
\cite{Teukolsky}). However, by interchanging $l$ with $n$ and $m$ with
$\bar{m}$ in what follows, one easily obtains the corresponding
results for $\Psi_0$, describing ingoing radiation at the event
horizon.

With respect to the chosen null tetrad, the pulsation operator acting
on the linearized field $\Psi^{(1)}_4$ splits into the sum of an
orbital and an angular operator,

\be
\left( \tilde{{\cal A}} + \frac{1}{r^2}\hat{{\cal A}} \right)
\Psi^{(1)}_4 = 0,
\label{Eq-Teukolsky}
\ee
where
\bea
\tilde{{\cal A}} &=& \left( \Delta + 2\gamma + 5\mu \right)
  \left( D + 4\epsilon - \rho \right) - 3\Psi_2, \nonumber\\
\hat{{\cal A}} &=& -\left( \bar{\hat{\delta}} - 2\hat{\alpha} \right)
          \left( \hat{\delta} + 4\bar{\hat{\alpha}} \right).
\label{Eq-AngOp}
\eea

Next, we compute the perturbed Weyl scalar $\Psi^{(1)}_4$: With
respect to the background metric (\ref{Eq-Metric}), one obtains

\bdm
\Psi^{(1)}_4 = R^{(1)}_{AbCd}\, n^b n^d \bar{m}^A\bar{m}^C.
\edm

Performing the multipole decomposition as described in section
\ref{Sect-2}, we obtain

\bdm
\Psi^{(1)}_4 = \left[ n^a n^b \nabtil_a h^{(inv)}_b \right]
  \left[ \bar{m}^A\bar{m}^B \nabhat_A S_B \right]
\edm
in the odd-parity sector and

\bdm
\Psi^{(1)}_4 = -\frac{1}{2} \left[ n^a n^b H^{(inv)}_{ab} \right]
  \left[ \bar{m}^A\bar{m}^B \nabhat_A\nabhat_B Y \right]
\edm
in the even-parity sector.  Using the definition of the derivative
operator $\hat{\delta}$ and the NP coefficient $\hat{\alpha}$, one can
check that in both parity sectors, the angular part is proportional to
the spin-weighted spherical harmonics $Y^{l m}_{-2}$ defined in
appendix \ref{App-B}.  Explicitly, we have
\bdm
Y^{l m}_{-2} = \frac{1}{C_l} \left( \bar{\hat{\delta}} - 2\hat{\alpha} \right)\bar{\hat{\delta}} Y^{l m},
\edm
where $C_l^2 = (l-1)l(l+1)(l+2)/4$.  It remains to express $h^{(inv)}_b$
and $H_{ab}^{(inv)}$ in terms of the scalar fields $\Phi$ and $\zeta$:
Using the definitions of the RW and Zerilli potentials $\Phi$ and
$\zeta$, as well as equations (\ref{Eq-ZerilliAmpl}) and
(\ref{Eq-LEQVTbis}), we obtain
\bea
h^{(inv)}_b &=& \tilde{\epsilon}_{ab}\nabtil^a ( r\Phi ), \nonumber\\
C_b &=& \frac{1}{a_0 + \lambda} \left[ -2N \zeta_{|b} - 2r^2 v^a
\zeta_{|a|b} - r^2 v_b\nabtil^a\nabtil_a\zeta \right], \nonumber
\eea
where we recall that $a_0 = 6M/r$.
Eventually, we get
\be
\Psi^{(1)}_4 = \sum\limits_{l m}
\left[ \frac{1}{r(a_0 + \lambda)} \left( \Delta + 2\gamma + 2\mu \right)\Delta (a_0 + \lambda)\Psi_{l m}
+ \frac{i}{r} \left( \Delta + 2\gamma + 2\mu \right)\Delta\Phi_{l m} \right] C_l Y^{l m}_{-2}\, ,
\label{Eq-ChanTrans}
\ee
which gives $\Psi^{(1)}_4$ in terms of the RW and Zerilli potentials
$\Phi_{l m}$ and $\Psi_{l m} = \zeta_{l m}/(a_0 + \lambda)$ introduced
in section 2. Here, we have re-introduced the indices $l$ and $m$.
In order for the metric perturbation to be real, we must have
$\bar{\Psi}_{l m} = \Psi_{l -m}$ and $\bar{\Phi}_{l m} = \Phi_{l -m}$
(a bar denoting complex conjugation).
Equation (\ref{Eq-ChanTrans}) (or its Fourier transform in time) is some
kind of generalization of the Chandrasekar transformation (see \cite{C-Book}, also
\cite{and_price}).

The corresponding expression for $\Psi^{(1)}_0$ follows after the
replacements $\Delta\mapsto D$, $\gamma\mapsto -\epsilon$, $\mu\mapsto
-\rho$ and $Y^{l m}_{-2}\mapsto Y^{l m}_2$ in the equation above.

Provided that $\Phi_{l m}$ and $\Psi_{l m}$ satisfy the RW and Zerilli
equations (\ref{Eq-RW2}) and (\ref{Eq-Zerilli}), respectively, and using
the vacuum NP equations and the commutation relations
\bdm
 \left( D + 2(s\!+\!1)\epsilon + q\,\rho \right) \left( \Delta + 2s\gamma + p\,\mu \right) -
  \left( \Delta + 2(s\!-\!1)\gamma + p\,\mu \right) \left( D + 2s\epsilon + q\,\rho \right)
= \frac{p + q}{2r^2} + 2(s + p + q)\Psi_2\, ,
\edm
where $s$, $p$ and $q$ are arbitrary real numbers, one can show that
indeed, $\Psi^{(1)}_4$ satisfies the Teukolsky equation
(\ref{Eq-Teukolsky}).

Finally, the total radiated energy per unit time can be obtained from
\be
\frac{ \diff E}{\diff u} = \lim\limits_{r\rightarrow\infty} \frac{r^2}{4\pi} \int_{S^2}
\left| \int\limits_{-\infty}^{u} \Psi_4(\tilde{u}, r, \Omega) \diff\tilde{u} \right|^2
\diff\Omega\, ,
\label{Eq-RadEnergy}
\ee
where asymptotically flat coordinates and an asymptotically flat NP
tetrad are chosen, and where $u = t-r$.  In our case $\Psi_4 = 0$ on
the background, so the radiated energy depends only quadratically on
$\Psi^{(1)}_4\,$.  Since the fields $\Phi_{l m}$ and $\Psi_{l m}$ are
scalars with respect to the background metric $\gtiltens$, we can evaluate
(\ref{Eq-RadEnergy}) using any asymptotically flat coordinates on the
background.  Using the fact that at infinity, $\Delta = \frac{1}{2}
(\p_t - \p_r) + O(r^{-1})$, $a,\gamma,\mu = O(r^{-1})$, and imposing
the outgoing wave condition $\dot{\Phi}_{l m} + \Phi'_{l m} = 0$,
$\dot{\Psi}_{l m} + \Psi'_{l m} = 0$ at infinity, one arrives at
\be
\frac{\diff E}{\diff u} = \frac{1}{16\pi}\lim\limits_{r\rightarrow\infty}
\sum\limits_{l \geq 2}\sum\limits_{m = -l}^l
\frac{ (l+2)! }{ (l-2)!} \left( |\dot{\Phi}_{l m}|^2 + |\dot{\Psi}_{l m}|^2 \right).
\label{energy}
\ee
(In the derivation, we have also used the orthogonality of the $Y^{l m}_{-2}$
and $\bar{\Psi}_{l m} = \Psi_{l -m}\,$, $\bar{\Phi}_{l m} = \Phi_{l-m}\,$.)
As a consistency check, it is useful to note that this coincides with
the usual well known result for Schwarzschild black holes in
Schwarzschild coordinates \footnote{Taking into account, of course,
that different normalizations are used in the literature when defining
the RW and Zerilli potentials, see, e.g. \cite{pl}.}.
Eq.(\ref{energy}), however, holds for any coordinates.

%%%%%%%%%%%%%%%%%%%%%%%%%%%%%%%%%%%%%%%%%%%%%%%%%%%%%%%%%%%%
\section{The RW and Zerilli equations and numerical issues}
%%%%%%%%%%%%%%%%%%%%%%%%%%%%%%%%%%%%%%%%%%%%%%%%%%%%%%%%%%%%

The RW and Zerilli equations are wave equations with exactly the same
differential operator (the Laplacian on the orbit space); they differ
only in the corresponding potentials, cf. eq. (\ref{master_lapla}).
This simplifies their analysis, since properties such as
well-posedness do not depend on lower order terms (as the potentials
are). We now discuss certain properties of the RW and Zerilli
equations. In particular, we wish to note the fact that they are
perfectly well defined as long as the background is regular (both in
the coordinate and curvature sense). Provided the latter holds there is
no pathology in the equations (or the solutions) at, for example, the
event horizon.

We start writing these equations explicitly by introducing
coordinates.  We then express the whole metric as $g^{total}_{\mu \nu}
= g_{\mu \nu} + \delta g_{\mu \nu}$, with the background metric given
by

\be
\gtens =
(-\alpha ^2+\gamma ^2\beta ^2)dt^2 + 2\gamma ^2\beta dt dx + \gamma ^2dx^2 +
r^2(d\vartheta ^2 + \sin ^2{\vartheta }d\phi ^2) \label{coordinate_metric}
\ee

The RW and Zerilli equations are

\begin{equation}
\ddot{Z} = c_1\dot{Z'}+  c_2 Z^{''} + c_3\dot{Z} + c_4Z'  - \alpha^2 V Z
\label{master}
\end{equation}

where $Z$ denotes either the RW or Zerilli functions. The coefficients
$c_i$ are

\begin{eqnarray*}
c_1 & = &  2 \beta  \\
c_2 &=&
 {\frac {\left (\alpha ^2 - \gamma ^2\beta ^2 \right
)}{\gamma ^2}}  \\
%%%%%%%%%%%%%%%%%%%%%%%%
c_3 & = & {\frac {\left (\gamma  \dot{\alpha} - \gamma  \beta
\alpha ' +\alpha \beta \gamma ' - \alpha \dot{\gamma } + \gamma  \alpha \beta '
\right )}{\gamma  \alpha  }} \\
c_4 &=&
{\frac{1}{\gamma ^3\alpha } \left (- \gamma ^3\beta \dot{\alpha } - \alpha ^3
\gamma ' + \gamma ^3 \beta ^2\alpha ' - 2 \gamma ^3\alpha \beta \beta ' +
\gamma ^3\alpha \dot{\beta } + \gamma ^2\alpha \beta \dot{\gamma } + \gamma
\alpha ^2\alpha ' - \gamma ^2 \alpha \beta ^2\gamma ' \right ) }
\end{eqnarray*}

and the corresponding potentials are

\begin{eqnarray*}
V_{RW} & = & \frac{1}{r^2}\left[ l(l+1) - \frac{6M}{r} \right], \\
V_{Z} & = & \frac{ \lambda^2 r^2 [ (\lambda+2)r + 6M) ] + 36 M^2 (\lambda r + 2M) }{(\lambda r + 6M)^2 r^3}\, .
\end{eqnarray*}

To make the hyperbolic character of these equations manifest, we
introduce new variables $y:=Z',w:=\dot{Z}$, and write it as a
first-order system for the ``vector'' $ u = (w,y,Z)^T$, i.e., $
\dot{u} = Au' + Bu$, where in this case the principal part is

$$
A = \left(
\begin{array}{ccc}
c_1 & c_2 &  0 \\
1 & 0 & 0 \\
0 & 0 & 0
\end{array}
\right)
$$

and has eigenvalues and eigenvectors

\begin{eqnarray*}
\lambda _0 &=& 0 \; , \; \mbox{ with } \vec{e}_0 = [0,0,1] \\
\lambda _{\pm} &=& \frac{1}{2}\left[c_1 \pm (c_1^2+4c_2)^{1/2} \right]
\; , \;  \mbox{ with } \vec{e}_{\pm} = [\lambda _{\pm},1,0]
\end{eqnarray*}

In our case, $c_1^2 + 4c_2 = 4\alpha ^2 \gamma ^{-2}$ and, so, the
 eigenvectors of $A$ are independent provided the background metric is
 locally well defined. Thus, the system is strongly hyperbolic, which
 is enough to prove well-posedness for the initial-boundary value
 problem if one gives boundary data for the characteristic modes that
 enter the domain \cite{kreiss_lorentz}. For the close-limit evolution
 of black holes in horizon-penetrating coordinates, for example, one
 would put the inner boundary inside the black hole, check that the
 characteristic modes are indeed leaving the computational domain
 (i.e. that the eigenvalues of $A$ are positive), and thus not put
 boundary conditions there (``excision''). At the outer boundary one
 would typically put zero boundary conditions for the ingoing modes.

One of the additional advantages of having a hyperbolic equation is
that one can write codes that can {\it a priori} be shown to be
convergent \cite{kreiss}. We have indeed written two such codes for
the RW and Zerilli equations with an arbitrary background.  One of
them uses fourth-order centered differences in space and fourth order
Runge-Kutta in time. It uses extrapolation at the inner boundary
(assumed to be inside the black hole), and gives zero boundary data to
the characteristic mode that enters the computational domain at the
outer boundary. The other code is second-order; it also uses
Runge-Kutta for time integration and centered differencing in space,
but now needs some dissipation (one can prove that this scheme is
unstable without dissipation, see \cite{kreiss}). In future work we
will present numerical details of these codes applied to the
close-limit collision of superposed Kerr-Schild and
Painlev\'e-Gullstrand black holes.

%%%%%%%%%%%%%%%%%%%%%%%%%%%%%%%%%%%%%%%%%%%%%%%%%%%%%%%%%%%%
\section{Acknowledgements}
%%%%%%%%%%%%%%%%%%%%%%%%%%%%%%%%%%%%%%%%%%%%%%%%%%%%%%%%%%%%
We would like to thank Abhay Ashtekar, Pablo Laguna and Jorge Pullin for
useful discussions, and Bernard Kelly for carefully reading the manuscript. This work was supported
by the Swiss National
 Science Foundation, by grants NSF-PHY-0090091, NSF-PHY-9800973, by Fundaci\'on Antorchas,
and by the Eberly Family Research Fund at Penn State.

\appendix

%%%%%%%%%%%%%%%%%%%%%%%%%%%%%%%%%%%%%%%%%%%%%%%%%%%%%%%%%%%%
\section{Perturbed four-metric in term of the potentials}
\label{App-A}
%%%%%%%%%%%%%%%%%%%%%%%%%%%%%%%%%%%%%%%%%%%%%%%%%%%%%%%%%%%%

Here we will give explicitly some of the expressions used in the body
of the paper. That is, we choose a general coordinate system for the
background metric, as in eq. (\ref{coordinate_metric}).

%%%%%%%%%%%%%%%%%%%%%%%%%%%%%%%%%%%%
\subsection{Odd-parity sector}
%%%%%%%%%%%%%%%%%%%%%%%

%%%%%%%%%%%%%%%%%%%%%%%
\subsubsection{Four metric}
%%%%%%%%%%%%%%%%%%%%%%%
The perturbation for the four metric with $l \geq 2$ is given by ($Y_{\phi} =
\partial _{\phi}Y$, etc.)
\begin{eqnarray*}
\delta g_{x\vartheta } & = & \left[\frac{\gamma }{\alpha }\left( -r\dot{\Phi} +
\beta r\Phi' + \Phi(\beta r' - \dot{r}) \right) + \frac{rk'-2kr'}{r}\right]
\frac{Y_{\phi}}{\sin{\vartheta }}  \\
&& \\
%%%%%%%%%%%%%%%%%
\delta g_{x\phi } & = & -\left[\frac{\gamma }{\alpha }\left( -r\dot{\Phi} + \beta r\Phi' +
\Phi(\beta r' - \dot{r}) \right) + \frac{rk'-2kr'}{r}\right]
 \sin{\vartheta }
Y_{\vartheta}  \\
&& \\
%%%%%%%%%%%%%%%%%%
\delta g_{\vartheta \vartheta } & = & \frac{2k}{\sin ^2{\vartheta }}\left[-\cos{\vartheta
}Y_{\phi}+\sin{\vartheta }Y_{\vartheta \phi } \right] \\
&& \\
%%%%%%%%%%%%%%%%%%
\delta g_{\vartheta \phi } & = & k\left[\cos{\vartheta }Y_{\vartheta }+\sin ^{-1}{\vartheta
}Y_{\phi \phi } -\sin{\vartheta}Y_{\vartheta \vartheta} \right] \\
&& \\
%%%%%%%%%%%%%%%%%%
\delta g_{\vartheta t } & = & \left[\frac{1}{\gamma \alpha }\left( -\gamma ^2\beta r\dot{\Phi}
 + r(\gamma ^2\beta ^2-\alpha ^2)\Phi' +
 (-\alpha ^2r'-\dot{r}\beta \gamma ^2+r'\gamma ^2\beta ^2 )\Phi \right)
 + \frac{r\dot{k}-2k\dot{r}}{r} \right]\frac{Y_{\phi}}{\sin{\vartheta }}  \\
& & \\
%%%%%%%%%%%%%%%%%%%
\delta g_{\phi \phi } & = & 2k \left[\cos{\vartheta
}Y_{\phi} - \sin{\vartheta }Y_{\vartheta \phi } \right] \\
&& \\
%%%%%%%%%%%%%%%%%%
\delta g_{\phi t } & = & -\left[\frac{1}{\gamma \alpha }\left( -\gamma ^2\beta r\dot{\Phi}
 + r(\gamma ^2\beta ^2-\alpha ^2)\Phi' +
 (-\alpha ^2r'-\dot{r}\beta \gamma ^2+r'\gamma ^2\beta ^2 )\Phi \right)
 + \frac{r\dot{k}-2k\dot{r}}{r} \right]\sin{\vartheta }Y_{\vartheta }
%%%%%%%%%%%%%%%%%%%
\end{eqnarray*}

It is straightforward to compute the linearized Ricci or Einstein
tensor for the above perturbed metric and see that they are indeed
annihilated if the master equation (\ref{master}) holds.

The $l=1$ components of the metric, on the other hand, are given by

\begin{eqnarray*}
%%%%%%%%%%%%%%%%%%%%%%%%%%%%%%%%%%%%%%%%
\delta g_{x \vartheta} &=& \left[\frac{f'r-2fr'}{r} +
\frac{2J\gamma (-\dot{r}+\beta r')}{(2M-r)\alpha}
\right]\frac{Y_{\phi}}{\sin{\vartheta }}\\
%%%%%%%%%%%%%%%%%%%%%%%%%%%%%%%%%%%%%%%%
& & \\
\delta g_{x \phi} &=& -\left[\frac{f'r-2fr'}{r} +
\frac{2J\gamma (-\dot{r}+\beta r')}{(2M-r)\alpha}
\right]\sin{\vartheta }Y_{\vartheta } \\
%%%%%%%%%%%%%%%%%%%%%%%%%%%%%%%%%%%%%%%%
& & \\
\delta g_{t \vartheta} &=& \left[\frac{\dot{f}r-2f\dot{r}}{r} +
\frac{2J\left (\gamma ^2\beta (\beta r'-\dot{r}) -\alpha ^2r'
\right)}{\gamma \alpha (2M-r)} \right]\frac{Y_{\phi}}{\sin{\vartheta }}\\
%%%%%%%%%%%%%%%%%%%%%%%%%%%%%%%%%%%%%%%%
& & \\
\delta g_{t \phi} &=& - \left[\frac{\dot{f}r-2f\dot{r}}{r} +
\frac{2J\left(\gamma ^2\beta (\beta r'-\dot{r}) -\alpha ^2r'
\right)}{\gamma \alpha (2M-r)} \right]\sin{\vartheta }Y_{\vartheta }
%%%%%%%%%%%%%%%%%%%%%%%%%%%%%%%%%%%%%%%%
\end{eqnarray*}

and it is also straightforward to check that this linearized metric
satisfies the linearized vacuum equations.

%%%%%%%%%%%%%%%%%%%%%%%%%%%%%%%%%%%%%%%
\subsubsection{Perturbed ADM quantities}
%%%%%%%%%%%%%%%%%%%%%%%%%%%%%%%%%%%%%%%

The three metric can be obtained straightforwardly from the spatial
components of the four metric explicitly given above and, similarly,
the coshift can be obtained from $\delta \beta _i = \delta g_{ti}$ and
the above expressions for $\delta g_{ti}$. On the other hand, the
nontrivial components of the perturbed extrinsic curvature can be
computed directly from the four-metric above explicitly, or from the
results in the body of the paper. In either case, the results for $l
\geq 2$, are

%%%%%%%%%%%%%%%%%%%%%%%%%%%%%%%%
\begin{eqnarray*}
%%%%%%%%%%%%%%%%%%%%%%%%%%%%%%%%
\delta K_{x\vartheta}& = &
\frac{Y_{\phi }}{r^2\alpha ^2\sin{\vartheta }}\left[\alpha (\dot{r}-\beta
r')(k'r-2kr') + \gamma r^2(-\dot{r}+\beta r')(\dot{\Phi}-\beta \Phi') +
\right. \\
& & \\
& & \left.  \frac{ar}{2}\left(-2(r')^2\beta ^2 + \alpha
^2(l(l+1)-2) - 2\dot{r}^2 + 4\beta \dot{r}r'\right)\Phi
\right]  \\
& & \\
%%%%%%%%%%%%%%%%%%%%%%%%%%%%%%%%
\delta K_{x\phi}& = & - \frac{Y_{\vartheta }\sin{\vartheta }}{r^2\alpha ^2}\left[\alpha (\dot{r}-\beta
r')(k'r-2kr') + \right. \\
& & \\
& & \left. \gamma r^2(-\dot{r}+\beta r')(\dot{\Phi}-\beta
\Phi') +  \frac{\gamma r}{2}\left(-2(r')^2\beta ^2 + \alpha ^2(l(l+1)-2) -
2\dot{r}^2 + 4\beta \dot{r}r'\right)\Phi
\right] \\
& & \\
%%%%%%%%%%%%%%%%%%%%%%%%%%%%%%%%
\delta K_{\vartheta \vartheta}& = & \left[ \frac{1}{\gamma }(r\Phi'+r'\Phi) +
\frac{2k(\dot{r}-\beta r')}{\alpha r} \right]\left(-\frac{Y_{\phi }\cos{\vartheta
}}{\sin ^2{\vartheta }} + \frac{Y_{\vartheta \phi }}{\sin{\vartheta}} \right) \\
& & \\
%%%%%%%%%%%%%%%%%%%%%%%%%%%%%%%%
\delta K_{\vartheta \phi}& = & \frac{1}{2}\left[ \frac{1}{\gamma }(r\Phi'+r'\Phi) +
\frac{2k(\dot{r}-\beta r')}{\alpha r} \right]\left(\frac{Y_{\phi \phi
}}{\sin{\vartheta }} + \cos{\vartheta}Y_{\vartheta}  - \sin{\vartheta}Y_{\vartheta \vartheta}
\right)  \\
& & \\
%%%%%%%%%%%%%%%%%%%%%%%%%%%%%%%%
\delta K_{\phi \phi}& = &\left[ \frac{1}{\gamma }(r\Phi'+r'\Phi) +
\frac{2k(\dot{r}-\beta r')}{\alpha r} \right]\left(Y_{\phi }\cos{\vartheta } -
\sin{\vartheta}Y_{\vartheta \phi } \right)
\end{eqnarray*}

and for $l=1$,

\begin{eqnarray*}
\delta K_{x\vartheta}& = & \left[\frac{(\dot{r}-\beta r')(f'r-2fr')}{\alpha r^2}-
 \frac{J}{r\gamma \alpha ^2 (r-2M)^2}
 \left( 4\gamma ^2M\beta \dot{b}b' - 2\gamma ^2M\alpha ^2 - 2\gamma
^2M\dot{r}^2 - 2\beta ^2\gamma ^2M (b')^2 + \right. \right.\\
& & \left. \left. r\gamma ^2 \alpha ^2 - 4\beta
\gamma ^2 b'b\dot{b} + 2\beta ^2 \gamma ^2 (b')^2b - 4\alpha ^2 (r')^2 r + 6
\alpha ^2 (r')^2M + 2(\dot{r})^2\gamma ^2 r
\right)\right]\frac{Y_{\phi}}{\sin{\vartheta}} \\
& & \\
%%%%%%%%%%%%%%%%%%%%%%%%%%%%%%%%
\delta K_{x\phi}& = & -\left[\frac{(\dot{r}-\beta r')(f'r-2fr')}{\alpha r^2}-
 \frac{J}{r\gamma \alpha ^2 (r-2M)^2}
 \left( 4\gamma ^2M\beta \dot{b}b' - 2\gamma ^2M\alpha ^2 - 2\gamma
^2M\dot{r}^2 - 2\beta ^2\gamma ^2M (b')^2 + \right. \right.\\
& & \left. \left. r\gamma ^2 \alpha ^2 - 4\beta
\gamma ^2 b'b\dot{b} + 2\beta ^2 \gamma ^2 (b')^2b - 4\alpha ^2 (r')^2 r + 6
\alpha ^2 (r')^2M + 2(\dot{r})^2\gamma ^2 r
\right)\right] Y_{\vartheta}\sin{\vartheta}
%%%%%%%%%%%%%%%%%%%%%%%%%%%%%%%%
\end{eqnarray*}

%%%%%%%%%%%%%%%%%%%%%%%%%%%%%%%%%%%%
\subsection{Even-parity sector}
%%%%%%%%%%%%%%%%%%%%%%%

The expressions in this sector are also straighforward to obtain, but the final expression are too
long to be written down here. For this reason we wil only present the simplest explicit example:
the perturbed four metric, in the RW gauge, for $l\geq 2$ perturbations of the Painlev\'e-Gullstrand
spacetime. The background metric is, thus, given by
$$
g_{xx}  =  1 \;\; , \;\;
g_{xt} = \left(\frac{2M}{r}\right)^{1/2}
g_{\vartheta \vartheta} =  r^2 \;\; , \;\;
g_{\phi \phi } = r^2 \sin ^2{\vartheta } \;\;
g_{tt}  =  -1 + \frac{2M}{r} \;\; \mbox{(where $r\equiv x$)} \; ,
$$
while the perturbation is
\begin{eqnarray*}
\delta g_{xx} & = & \left[ 3\left( \frac{2M}{r}\right ) ^{1/2}\left( \frac{rl(l+1) - 2r
+2M}{6M-2r+rl(l+1)}\right) \dot{\Psi } - 2r\Psi ^{''} -
\left( 2\frac{r(r+3M)l(l+1)-2r^2-6M(r-M)}{r(6M-2r+rl(l+1))} \right) \Psi ' \right.  \\
& & \\
%%%%%%%%%%%%%%%%%%%%
& & + \left( 3r^3l^5 + r^3l^6 - 7l^3r^3 - l^4r^3 + 4lr^3 - 18Mr^2l^2
+ 12Mr^2l^3 + 6Mr^2l^4 - 24Mr^2(l-1) + 36M^2rl(l+1) + \right.  \\
& & \\
%%%%%%%%%%%%%%%%%%
& &  \left.  \frac{ \left.
 72M^2(M-r)\right)}{\left( 6M-2r+rl(l+1)\right )^2r^2} \Psi  \right]Y \\
& & \\
%%%%%%%%%%%%%%%%%%%%
\delta g_{xt}&=& \left[ -2r\dot{\Psi '} -2\frac{r(r-3M)l(l+1) - 2r^2 +
6M(r-M)}{r^{3/2}(6M-2r+rl(l+1))}\left( r^{1/2}\dot{\Psi } - (2M)^{1/2}\Psi ' \right) +
\right. \\
& & \\
%%%%%%%%%%%%%%
& &  (2M)^{1/2}\left( 3r^3l^5 + r^3l^6 - 7l^3r^3 - l^4r^3 + 4lr^3 - 18Mr^2l^2
+ 12Mr^2l^3 + 6Mr^2l^4 - 24Mr^2(l-1) +  \right. \\
& & \left. \frac{\left. 36M^2rl(l+1) +  72M^2(M-r) \right)}
{\left(6M-2r+rl(l+1)\right)^2 r^{5/2}} \Psi \right]Y  \\
%%%%%%%%%%%%%%%%%%%
\delta g_{\vartheta \vartheta } & = & \left[ -(2r)^{3/2}M^{1/2}\dot{\Psi } - 2(r-2M)r\Psi '
- \left( \frac{-12M(r-2M) + 6Mrl(l+1) - r^2l^2 + 2r^2l^3 + r^2l^4 - 2r^2l}{6M-2r+rl(l+1)}\right)
\Psi \right]Y \\ & & \\
%%%%%%%%%%%%%%%%%%%
\delta g_{\phi \phi }& = & \left[ -(2r)^{3/2}M^{1/2}\dot{\Psi } - 2(r-2M)r\Psi '
- \left( \frac{-12M(r-2M) + 6Mrl(l+1) - r^2l^2 + 2r^2l^3 + r^2l^4 -
 2r^2l}{6M-2r+rl(l+1)}\right) \Psi \right] \times \\
& & Y\sin ^2{\vartheta } \\
& & \\
%%%%%%%%%%%%%%%%%%%
\delta g_{tt} &=& \left[-4(2rM)^{1/2}\dot{\Psi '} -
\left( \frac{(2M)^{1/2}(r^2l^2 - 6Mr(l^2+l-1) + r^2l - 2r^2 -12M^2)}
{r^{3/2}(6M-2r+rl(l+1))} \right) \dot{\Psi } - 2(r-2M)  \Psi ^{''} - \right. \\
&& \\
& & \left( 2\frac{r(r^2-3Mr+6M^2)l(l+1) - 2r^3 + 6Mr(r-M) + 12M^3 }{r^2(6M-2r+rl(l+1))}
\right) \Psi ' +  (r+2M) \times \\
& & \\
& & \left( 3r^3l^5 + r^3l^6 - 7l^3r^3 - l^4r^3 + 4lr^3 - 18Mr^2l^2
+ 12Mr^2l^3 + 6Mr^2l^4 - 24Mr^2(l-1) + 36M^2rl(l+1) +  \right. \\
& & \left. \frac{\left. 72M^2(M-r) \right)} {\left(6M-2r+rl(l+1)\right)^2 r^3 } \Psi
%%%%%%%
\right]Y
\end{eqnarray*}

%%%%%%%%%%%%%%%%%%%%%%%%%%%%%%%%
\section{Spin-weighted spherical harmonics}
%%%%%%%%%%%%%%%%%%%%%%%%%%%%%%%%
\label{App-B}

We define the operators $c(s)$ and their adjoints $c^\dagger(s)$ by
\bdm c(s) = -\left( \bar{\hat{\delta}} + 2(s+1)\hat{\alpha} \right),
\;\;\; c^\dagger(s) = \left( \hat{\delta} - 2s\bar{\hat{\alpha}}
\right), \edm respectively.  The angular operator defined in
(\ref{Eq-AngOp}) takes the form $\hat{{\cal A}} = c(-2)c^\dagger(-2)$.
Using the commutation relations
\bdm
c(s) c^\dagger(s) - c^\dagger(s-1) c(s-1) = -s
\edm
for any real number $s$ we can construct the eigenfunctions of
$\hat{{\cal A}}_s = c(s)c^\dagger(s)$ from the the standard spherical
harmonics $Y^{l m}$, which fulfill
\bdm
c(0) c^\dagger(0) Y^{l m} = -\frac{1}{2}\hat{\Delta} Y^{l m} = \frac{1}{2}l(l+1) Y^{l m}.
\edm
The eigenfunctions of $\hat{{\cal A}}_s$ are called the {\it
spin-weighted spherical harmonics}, and are proportional to
\bdm
Y^{l m}_s \equiv \left\{ \begin{array}{l}
\frac{1}{C_{ls}} c^\dagger(s-1) c^\dagger(s-2)\cdot\cdot\cdot c^\dagger(0) Y^{l m},\;\;\; s \geq 0,  \\ \\
\frac{1}{C_{l-s}} c(s)c(s+1)\cdot\cdot\cdot c(-1) Y^{l m},\;\;\; s < 0
\end{array} \right. ,
\edm
where
\bdm
C_{ls}^2 = \frac{1}{2^s}(l-s+1)(l-s+2)\cdot\cdot\cdot(l+s)
\edm
is a normalization constant.
For all $s$, we have
\bdm
\hat{{\cal A}}_s Y^{l m}_s = \frac{1}{2}(l-s)(l+s+1) Y^{l m}_s\, .
\edm

%%%%%%%%%%%%%%%%%%%%%%%%%%%%%%%%%%%%%%%%%%%%%%%%

\end{document}